\documentclass{pos}

\title{CMS: Cosmic muons in simulation and measured data}

\ShortTitle{CMS: Cosmic muons}

\author{\speaker{Lars Sonnenschein} 
     \\
     RWTH Aachen University, III. Physikalisches Institut A, 52056 Aachen, Germany \\
     E-mail: \email{Lars.Sonnenschein@cern.ch} \\
     {\bf on behalf of the CMS collaboration} \\     
   Conference report CMS CR-2010/006, submitted to proceedings of Hadron Collider Physics 09
}

\abstract{
A dedicated cosmic muon Monte-Carlo event generator CMSCGEN has been 
developed for the CMS experiment.
The simulation relies on parameterisations of the muon energy and the
incidence angle, based on measured and simulated data of the cosmic muon flux.
The geometry and material density of the CMS infrastructure underground 
and surrounding geological layers are also taken into account.
The event generator is integrated into the CMS detector simulation chain of the 
existing software framework.
Cosmic muons can be generated on earth's surface as well as for the detector 
located 90~m underground.
Many million cosmic muon events have been generated and compared to measured data,
taken with the CMS detector at its nominal magnetic field of 3.8 T.
}

\FullConference{XXth Hadron Collider Physics Symposium \\
		 November 16 -- 20, 2009 \\
		 Evian, France}

\begin{document}

\section{Muon flux parameterisation and normalisation}

\noindent
The cosmic muon generator 
of the CMS experiment~\cite{cms} is based on a parameterised differential muon flux
obtained by means of the air shower program CORSIKA~\cite{corsika}
which in turn has been operated making use of the EPOS interaction model 
for high energy interactions and the
GHEISHA interaction model for low energy hadronic interactions.
The CORSIKA simulation results have been fitted with polynomials~\cite{cmscgen}
to describe the differential flux at earth's surface as a function 
$\frac{d\Phi}{dp\; d{\cos\theta}\; d\phi}$ 
of the azimuth angle $\phi$, the incidence angle $\theta$ and the muon momentum $p$.
The flux is normalised to vertical muons from measured data with an energy of 100~GeV.
The maximally allowed phase space in azimuth angle is
$\phi\in[0,2\pi)$, whereas azimuthal isotropy has been assumed.
The phase space for incidence angles is given by $0^{\circ}<\theta<84^{\circ}$.
For incidence angles of $\theta>75^{\circ}$ the parameterisation is extrapolated.
In momentum the phase space is restricted to $3<p_{\mu}<3000$~GeV.
The lower limit is driven by the existence of new physical processes which set in 
at such low momenta which makes the extrapolation invalid. 
The upper limit of muon momenta is driven by strongly decreased flux in this regime. 
The muon momentum is approximated by a polynomial of the expression  
\begin{equation}
    L=\log_{10}(p/\mbox{GeV}) 
\end{equation}
which is slowly varying in muon momentum.
The muon momentum spectrum is then fitted  by a polynomial of the form 
\begin{equation}
    s(L)=a_0+a_1L+...+a_6L^6 .
\end{equation}
The momentum dependent zenith angle is taken into account by a polynomial
of the cosine of the incident angle 
\begin{equation}
  z(\cos\theta,L)=b_0(L)+b_1(L)\cos\theta+b_2(L)\cos^2\theta .
\end{equation}
Finally the differential flux is given by  
\begin{equation}
\frac{d\Phi}{dp\; d{\cos\theta}\; d\phi}=C_{\mbox{\scriptsize norm}}\cdot \frac{1}{p^3}\cdot s(L) \cdot z(\cos\theta,L)\cdot\frac{1}{2\pi} .
\end{equation}

\section{Muon energy loss in material}

\noindent
A random vertex on a disk at the surface is assigned to a generated single muon.
The size of the disk which is centred around the vertical symmetry axis of the detector
is determined by the chosen maximal incidence angle such, that
a muon whose extrapolated direction does hit the CMS detector does necessarily have a
vertex inside the disk.
In the case the muon direction is pointing to the CMS detector the muon is propagated
through the different material densities of the geological environment between the surface
and the CMS detector, located at 90~m underground
to obtain an integrated amount of water equivalents. 
This amount determines the energy loss of a given muon as a 
parameterised function~\cite{pdg04}
of the energy of the incident muon at surface.
If the direction is not pointing to the CMS detector a new random vertex at the surface is 
chosen.
A material map describes the diverse materials from earth's surface
to the CMS detector.
In concrete, the fundament of the hall at surface, 
the three vertical access shafts, a movable plug of the main shaft 
as well as 
the collision and service caverns including the adjacent parts of the LHC tunnel 
are taken into account. Two different average densities
are assigned to the geological layers surrounding the CMS infrastructure. The upper half
consists of sand, clay, gravels and water while the lower half consists of rock.
In Fig. \ref{VxzSurface&CRAFT} the vertex position of the muons at the surface which
reach the CMS detector are shown. 
Lower energetic muons are responsible for the enhanced intensities
at the vertical access shafts. 
After the generated muons have been propagated to the CMS detector a full 
GEANT~\cite{geant} simulation of the detector processes the muons further.

\section{Simulation and data comparison}

\noindent
In the cosmic data taking period 2008 three hundred million events have been recorded with a
solenoid magnetic field of $B=3.8$~T. 
%
\begin{figure}[t]
 \unitlength 1cm
  \begin{picture}(11.0,4.6)

    \put(-0.5, -0.3){\includegraphics[width=.54\linewidth]{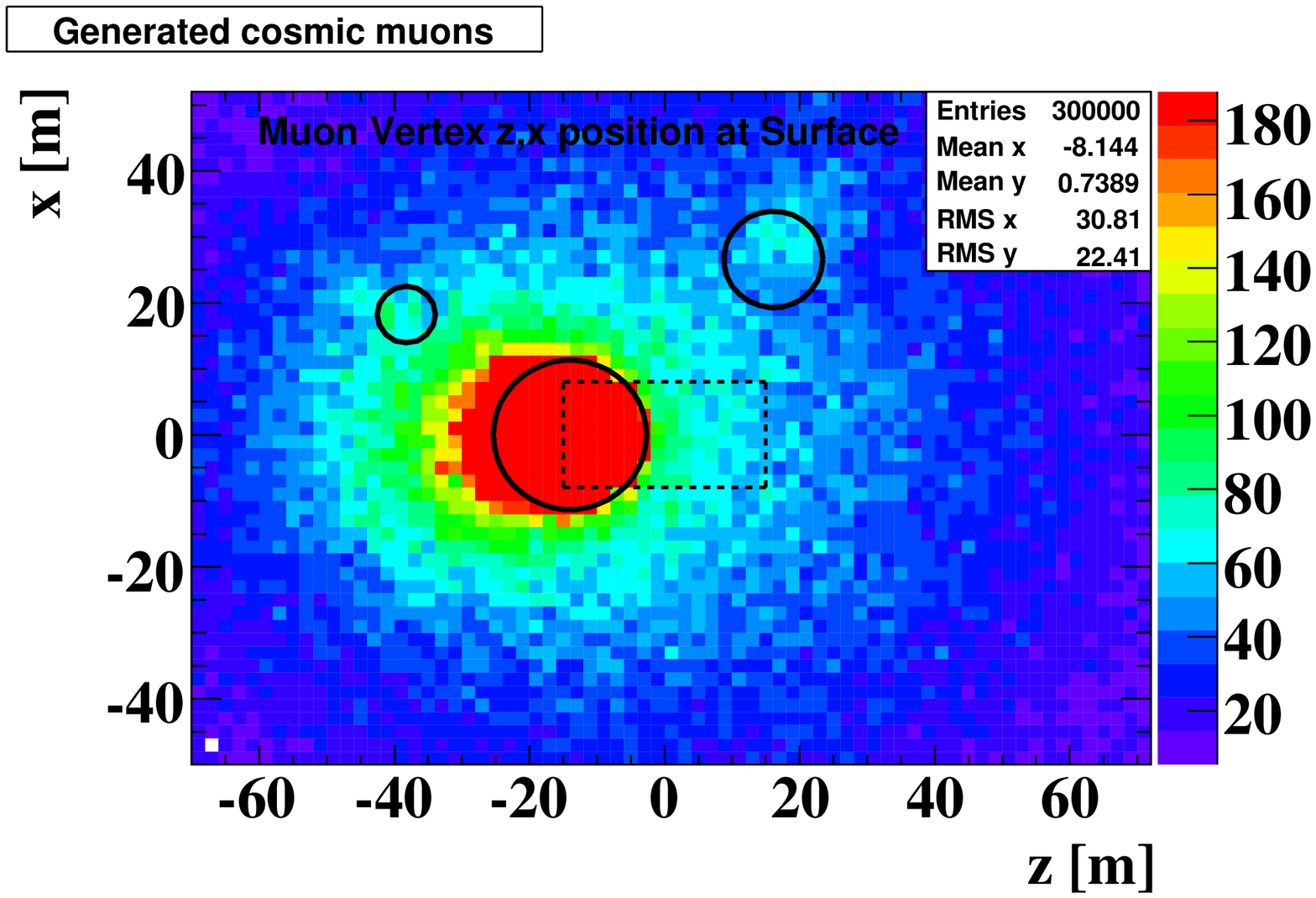}}

    \put(7.7, -0.4){\includegraphics[width=.55\linewidth]{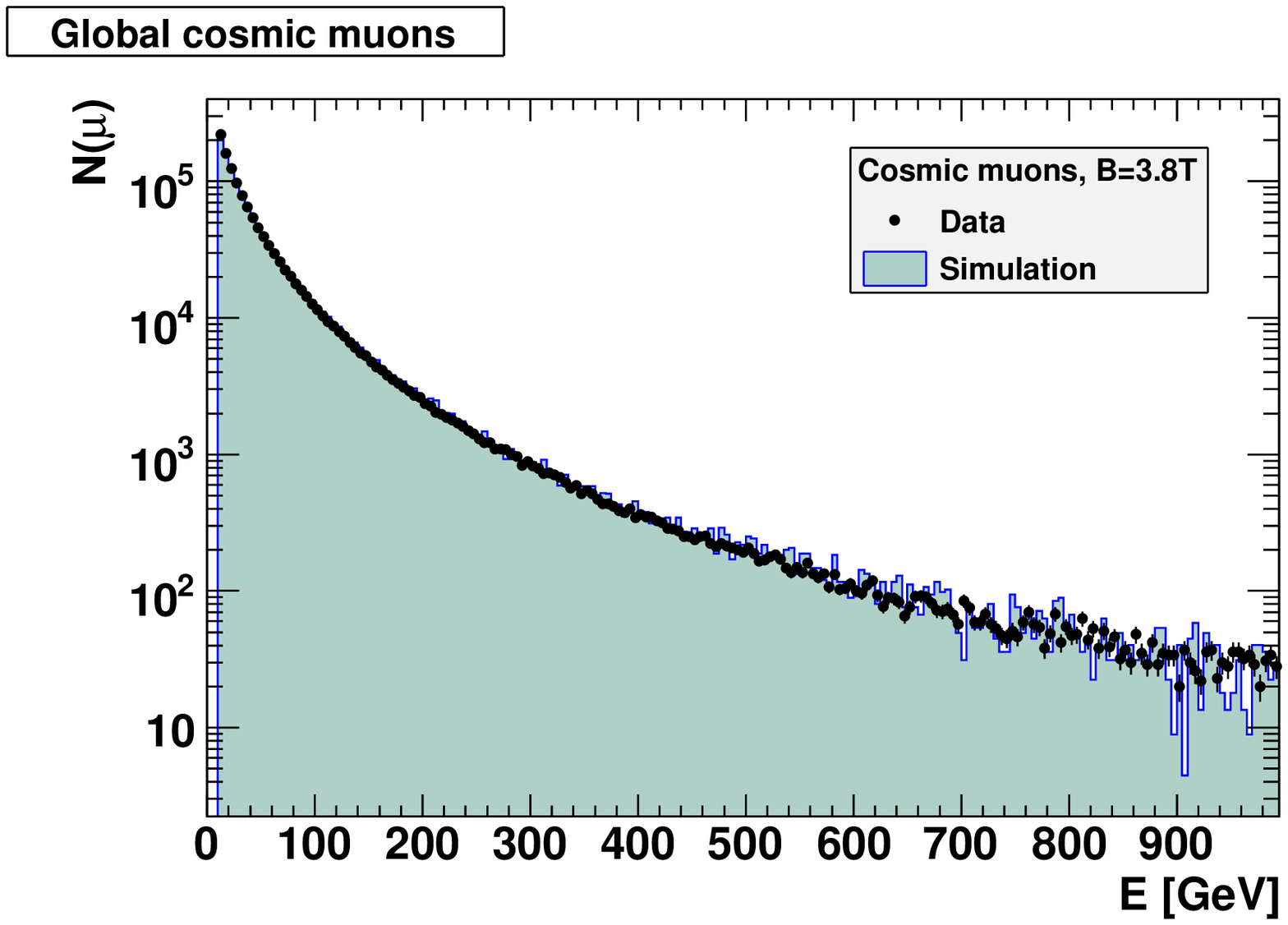}}
    \put(11.8, 4.85){CMS preliminary}

  \end{picture}

\caption{ \label{VxzSurface&CRAFT}
Left  plot: Entrance vertex at the surface of the generated muons which arrive at the
CMS detector (indicated by the dashed rectangle) 90~m underground. The three vertical
shafts are indicated by black circles. There is clearly a correlation between muon intensity
and shafts visible.
Right plot: Energy spectrum of muons in data (black points) measured with the CMS detector.
In comparison the GEANT simulation (green/blue histogram) is shown. 
The simulation has been normalised to the number of events of the data.
}
\vspace*{-1ex}
\end{figure}
%
%
%
Global cosmic muons, which are reconstructed
in the muon chambers and the central tracking system are chosen. A mixture of four different
triggers is used in the simulation. The momentum of the reconstructed
global muons is required to exceed a momentum threshold of $p=10$~GeV. The incidence angle
is restricted to the interval $0\leq \theta < 60^{\circ}$ to ensure that the parametrisation 
is applied 
far away from its regime of extrapolation.
Fig. \ref{VxzSurface&CRAFT} right plot shows the energy spectrum of the cosmic muon data taken
in 2008 in black points. The simulation is superposed as a green/blue shaded histogram and normalised
to the number of entries in data. 
The decrease of the data distribution by the power of -2.7 is very well described by the simulation.

\end{document}